\documentclass[twocolumn,preprintnumbers,amsmath,amssymb,superscriptaddress]{revtex4}

\usepackage{color}
\usepackage{blindtext}
\usepackage{graphicx}
\usepackage{dcolumn}
\usepackage{bm}
\usepackage{gensymb}
\usepackage{animate}

\def\be{\begin{equation}}
\def\ee{\end{equation}}
\def\bea{\begin{equation*}}
\def\eea{\end{equation*}}
\def\bna{\begin{eqnarray*}}
\def\ena{\end{eqnarray*}}
\def\bn{\begin{eqnarray}}
\def\en{\end{eqnarray}}
\def\bpm{\begin{pmatrix}}
\def\epm{\end{pmatrix}}

\def\be{\begin{equation}}
\def\ee{\end{equation}}
\def\bea{\begin{eqnarray*}}
\def\eea{\end{eqnarray*}}

\newcommand{\bra}[1]{\langle#1|}

\newcommand{\ket}[1]{|#1\rangle}

\newcommand{\braket}[1]{\langle#1\rangle}

\begin{document}

\title{Extracting the internal nonlocality from the dilated Hermiticity}

 \author{Minyi Huang}
 \email{hmy@zstu.edu.cn}
 \affiliation{Department of Mathematical Sciences, Zhejiang Sci-Tech University, Hangzhou 310007, PR~China}

\author{Ray-Kuang Lee}
 \email{rklee@ee.nthu.edu.tw}
\affiliation{Institute of Photonics Technologies, National Tsing Hua University, Hsinchu 300, Taiwan}
\affiliation{Center for Quantum Technology, Hsinchu 30013, Taiwan}
\affiliation{Physics Division, National Center for Theoretical Sciences, Taipei 10617, Taiwan}
 \author{Junde Wu}
 \email{wjd@zju.edu.cn}
\affiliation{School of Mathematical Sciences, Zhejiang University, Hangzhou 310027, PR~China}

\begin{abstract}
To effectively realize a $\cal PT$-symmetric system, one can dilate a $\cal PT$-symmetric Hamiltonian to some global Hermitian one and simulate its evolution in the dilated Hermitian system.
However, with only a global Hermitian Hamiltonian, how do we know whether it is a dilation and is useful for simulation?
To answer this question, we consider the problem of how to extract the internal nonlocality in the Hermitian dilation. We unveil that the internal nonlocality brings nontrivial correlations between the subsystems.
By evaluating the correlations with local measurements in three different pictures, the resulting different expectations of the Bell operator reveal the distinction of the internal nonlocality.
When the simulated $\cal PT$-symmetric Hamiltonian approaches its exceptional point, such a distinction tends to be most significant.
Our results clearly make a difference between the Hermitian dilation and other global Hamiltonians without internal nonlocality.
They also provide the figure of merit to test the reliability of the simulation, as well as to verify a $\cal PT$-symmetric (sub)system.

\end{abstract}

\maketitle

\section{Introduction}
Nowadays, both in classical and quantum physics, we are witnessing a growing interest in discussing
$\cal PT$-symmetric systems.
Historically, parity-time ($\cal PT$)-symmetric systems were first introduced to permit
entirely real spectra even in the non-Hermitian setting~\cite{bender1998real}.
With the experimental controls in gain and loss, photonic systems have been used to simulate $\cal PT$-symmetric wave
phenomena with an equivalence between single-particle quantum mechanics and classical wave
equation~\cite{El-OL, Makris-PRL, Guo-PRL, Ruter-NP, Chang-NP}.
Interestingly, $\cal PT$-symmetry has also found powerful applications in circuits design~\cite{Assaw-Nat}.
The concept of $\cal PT$-symmetry was generalized to the pseudo-Hermiticity~\cite{mostafazadeh2002Pseudo1,mostafazadeh2002Pseudo2,mostafazadeh2002Pseudo3,mostafazadeh2010pseudo},
and anti-$\cal PT$-symmetry~\cite{anti-1, anti-2}. Moreover, $\cal PT$-symmetric theory extended profoundly
to the researches of non-Hermiticity, with fruitful results and novel phenomena~\cite{ashida2020nonhermitian}.
For example, in the field of dynamics and  band topology,  the skin effect was introduced in the complex spectra
of non-Hermitian systems~\cite{PhysRevLett.121.086803,PhysRevLett.121.136802,PhysRevLett.123.170401,PhysRevResearch.1.023013}.

Despite the initial motivation to establish an alternative  framework of quantum theory, we can also
take $\cal PT$-symmetric systems as effective descriptions of large Hermitian systems in some subspaces, similar
to the Feshbach formalism dealing with an effective description of open quantum systems~\cite{Feshbach}.
By using the Naimark dilation theorem,  one can always find some four dimensional Hermitian Hamiltonians to effectively realize two dimensional unbroken $\cal PT$-symmetric systems~\cite{gunther2008naimark}.
Then, such a methodology can be generalized to simulate any finite dimensional $\cal PT$-symmetric systems~\cite{huang2017embedding,PhysRevLett.119.190401}.
 By evolving states under the Hermitian dilation Hamiltonians, and projecting out the
 ancillary systems, this paradigm successfully simulates the evolution of unbroken $\cal PT$-symmetric Hamiltonians in
 subspaces. It endows a direct physical
 meaning of $\cal PT$-symmetric  quantum systems in the sense of open systems.
As for the broken $\cal PT$-symmetry, there are at least two different approaches to the simulation. One way is
utilizing weak measurement as an approximation paradigm for the broken $\cal PT$-symmetric systems \cite{PhysRevLett.123.080404}; while the other way
is simulating the evolution of broken $\cal
PT$-symmetric systems with the time dependent Hamiltonians, connecting the topology and dynamics \cite{Wu878}.

In the simulation of $\cal PT$-symmetric systems, Hermitian dilation Hamiltonians play a key role.
Generally, these Hamiltonians are inseparable and act on certain global systems composed of two subsystems~\cite{huang2017embedding,PhysRevLett.119.190401}.
This implies that there exist non-local correlations between the subsystems.
To investigate such correlations, we propose an approach to extract the internal nonlocality when the global Hermitian Hamiltonian is shared with Alice and Bob.
With only local measurements performed by Alice and Bob, we show that the expectations of the Bell operator differ in different correlation pictures.
It gives a higher value of the upper bound when both the classical and local Hermitian pictures are considered, but a lower value of the upper bound for the simulation picture.
Moreover, when the simulated $\cal PT$-symmetric Hamiltonian approaches its exceptional point, the value of this upper bound gives the largest departure from the local Hermitian systems.
Along the line of quantum information approaches ~\cite{Lee14, Tang-NP,
Huang_2018,varma2019temporal,naikoo2019maximal}, our work provides a way to  distinguish isospectral Hermitian Hamiltonians with and without internal nonlocality.
With the ability to know whether a $\cal PT$-symmetric subsystem is embedded inside a global Hermitian one, our results provide the figure of merit to test the reliability of the simulation, as well as the verification for  a $\cal PT$-symmetric
(sub)system.

The remainder of this paper is organized as follows. In Sec. II, we introduce the preliminaries on the related notions
of $\cal PT$-symmetric systems and the CHSH (Clauser, Horne, Shimony, and Holt) scenario.
In Sec. III, the internal nonlocality is illustrated by investigating the correlation measurements between Alice and Bob, for the classical, local Hermitian and simulation picture, respectively.
Discussions on the physical implications of these three pictures and their
essential difference from the CHSH scenario are given in Sec. IV. Finally, in Sec. V we
conclude our results.

\section{Preliminaries}

\subsection{Basic notions and the simulation by dilation}
The concepts of parity, time reversal and $\cal PT$-symmetric operators have been studied since the early age of quantum mechanics.
A linear operator $H$ is said to be $\cal PT$-symmetric if $H{\cal PT}={\cal PT}H$, where $\cal P$ and $\cal T$ are
parity and time reversal operators.
Here, we  focus on finite dimensional spaces, in which a $\cal PT$-symmetric operator $H$ is said to be unbroken if it is
similar to a real diagonal operator; $H$ is said to be broken $\cal PT$-symmetric if it cannot be diagonalized or has complex eigenvalues \cite{huang2017embedding}.
$\cal PT$-symmetric operators are usually non-Hermitian but satisfy the condition of pseudo-Hermiticity \cite{mostafazadeh2010pseudo}.

The simulation of $\cal PT$-symmetric systems is closely related to the mathematical concept of Hermitian dilation~\cite{gunther2008naimark}.
Let $H$ be a $\cal PT$-symmetric operator on $\mathbb C^n$, and let $\hat{H}$ be a Hermitian operator on $\mathbb{C}^m$, with
$m>n$. $P_1$ is an operator defined by
$P_1:\mathbb{C}^m\rightarrow\mathbb{C}^n$, $P_1
\begin{bmatrix}\begin{smallmatrix}\phi_1\\\phi_2\end{smallmatrix}\end{bmatrix}=\phi_1$, where $\phi_1\in \mathbb{C}^n$ and $\phi_2\in
\mathbb{C}^{m-n}$.
Let
$X_{\hat{H}}=\{x : x\in \mathbb C^m, P_1\hat{H}
x=H P_1x,  P_1 e^{-it\hat{H}}x=e^{-itH} P_1 x
\}.$
If $ P_1X_{\hat{H}}=\mathbb{C}^n$, then we say that $\hat{H}$ is a Hermitian
dilation of $H$.

By evolving the  Hermitian dilation Hamiltonian $\hat{H}$ on a large space, the  evolution of a $\cal
PT$-symmetric non-Hermitian system can be realized in the subspace.
In addition, only unbroken $\cal PT$-symmetric operators preserve such a property, hence our discussions mainly focus on the case of unbroken $\cal PT$-symmetry.
Actually, for any state $\ket{\psi}$, the definition of $\hat{H}$ can ensure the following equation (unnormalized for convenience),
\bn
&&e^{-it\hat{H}}(\ket{0}\ket{\psi}+\ket{1}\ket{\tau\psi})=\ket{0}\ket{e^{-itH}\psi}+\ket{1}\ket{\tau
e^{-itH}\psi}\label{s2}, \nonumber
\en
where $\tau$ is an operator linked to the metric operator \cite{huang2017embedding}.
The equation above clearly shows that there are two subsystems.
Moreover, by projecting out $\ket{1}\ket{\tau
e^{-itH}\psi}$, the effect of a non-Hermitian Hamiltonian can be realized by $\ket{0}\ket{e^{-itH}\psi}$ in the latter subsystem.
An experimental realization of such an evolution effect in the subsystem is called a simulation, in
which the preparation of Hermitian dilation Hamiltonian plays an important role. As for more technical and experimental details, see \cite{Tang-NP,huang2017embedding}.

In general, there exist various $\hat{H}$ to satisfy the dilation scenario.
Without loss of generality, we adopt the following one by requiring
\bn
&&e^{-it\hat{H}}(\ket{1}\ket{\psi}-\ket{0}\ket{\tau \psi})=\ket{1}\ket{e^{-itH}\psi}-\ket{0}\ket{\tau e^{-itH}
\psi}\label{s4}.\nonumber
\en
Now, the resulting $\hat{H}$ can achieve a simple form,
\bn
\hat{H}&=&I_2\otimes \Lambda+i\sigma_y\otimes \Omega,\label{Ht1}\\
\Lambda&=&(H\tau^{-\frac{1}{2}}+\tau^{\frac{1}{2}}H)(\tau^{-\frac{1}{2}}+\tau^{\frac{1}{2}})^{-1},\\
\Omega&=&(H-\tau^{\frac{1}{2}}H\tau^{-\frac{1}{2}})(\tau^{-\frac{1}{2}}+\tau^{\frac{1}{2}})^{-1},\label{omega1}
\en
in which the details about $\tau$ can be referred to Refs. \cite{gunther2008naimark,huang2017embedding,PhysRevLett.119.190401}.

According to Eq. (\ref{Ht1}), the Hermitian dilation Hamiltonian $\hat{H}$ is inseparable.
That is, $\hat{H}$ cannot be written as a tensor product of two local operators.
As a consequence, it implies the possibility of investigating the correlations that $\hat{H}$ brings to the subsystems.
Moreover, $\hat{H}$ is isospectral to $H$.
It means that the Hermitian dilation Hamiltonian has the same eigenvalues as
the simulated $\cal PT$-symmetric Hamiltonian with twofold spectra, i.e., it has two multiplicities of eigenvalues.
Such a property implicitly allows us to use the measurements on the large space to simulate the
measurements of the $\cal PT$-symmetric system~\cite{PhysRevLett.123.080404}.
Briefly speaking, by measuring the Hermitian dilation Hamiltonian $\hat{H}$, one can read out the eigenvalues of the $\cal PT$-symmetric system.

\subsection{Two dimensional model}
To illustrate our proposed concept in a clear way, we consider a two dimensional $\cal PT$-symmetric Hamiltonian as an example by following ~\cite{gunther2008naimark,bender2007making}, i.e.,
\be H=E_0I_2+s\begin{bmatrix} i\sin\alpha&1\\1&-i\sin\alpha\end{bmatrix}.\label{HG}\ee
The corresponding eigenvalues for this two dimensional non-Hermitian system are $\lambda_\pm=E_0\pm s\cos\alpha$.
Moreover, there exists an exceptional point when $\alpha=\frac{\pi}{2}$,  in which case the Hamiltonian is no longer diagonalized. When $\alpha \neq \frac{\pi}{2}$,  the Hamiltonian $H$ has real eigenvalues and can be diagonalized. Hence, $\cal
PT$-symmetry is unbroken.
Specifically, when $\alpha=0$, the Hamiltonian returns to the Hermitian one.

By applying the dilation process given in Eqs. (\ref{Ht1}-\ref{omega1}), the corresponding Hermitian dilation Hamiltonian $\hat{H}$ has the form
\bn
\hat{H}&=&I_2\otimes \Lambda+i\sigma_y\otimes \Omega,\label{Ht}\\
\Lambda&=&E_0I_2+\frac{\omega_0}{2}\cos\alpha \sigma_x,\\
\Omega&=&i\frac{\omega_0}{2}\sin\alpha\sigma_z,\label{omega}
\en
where $\omega_0=2s\cos\alpha$~\cite{gunther2008naimark,huang2017embedding}.

\subsection{CHSH scenario}
As the joint correlation measurements will be performed locally by Alice and Bob, it is instructive to briefly recall Bell's nonlocality and the related CHSH scenario~\cite{Bellnonlocality,CHSH}.
In a standard Bell's test on nonlocality, two (sub)systems shared by Alice and Bob are spatially separated.
By performing local measurements, Alice obtains several possible outcomes from her subsystem, denoted as $a$, with the outcomes denoted as $b$ from  Bob's measurements on his subsystem.
Due to the randomness in the local measurements, the outcomes $a$ and $b$ may have different values. Nevertheless, these outcomes are in general governed by a probability distribution $p(ab|ij)$, where the local measurements are labeled with the index $i, j$.
Usually the joint probability distribution reveals that
\be
p(ab|ij)\neq p(a|i)p(b|j).\label{bell1}
\ee
It implies that the results $a$ and $b$ are not
independent, even when Alice and Bob are spacelike separated.
However, a classical correlation theory does not admit
nonlocality. Hence, a possible explanation is that some dependence between the subsystems was  established when they interacted in the past, eventually leading to the inequality shown in Eq. (\ref{bell1}).
Such an explanation also suggests that if we take into account all the past factors, described by some random variable $\nu$, then the joint probability distribution for $a$ and $b$ can be factorized as
\be
p(ab|ij,\nu)= p(a|i,\nu)p(b|j,\nu).\label{bell2}
\ee
Apparently, Eq. (\ref{bell2}) shows that $a$ and $b$ are independent, which is consistent with the classical (local) correlation theory.
On the other hand, by denoting $q(\nu)$ as the probability distribution of $\nu$, one can have
\be
p(ab|ij)=\int_N d\nu q(\nu)p(a|i,\nu)p(b|j,\nu),\label{bell3}
\ee
which is the condition for locality in the context of Bell's test.

Equation (\ref{bell3}) is the key to deriving the well-known CHSH inequality.
Suppose Alice can perform two local measurements denoted as $A_i, i\in\{0,1\}$; while Bob can also perform two local measurements $B_j, j\in\{0,1\}$. The possible outcomes of $A_i$ and $B_j$ have two values labeled $a, b\in \{+1, -1\}$.
Now, let $\braket{A_i B_j}=\sum_{a,b}ab\,p(ab|ij)$ be the expectation value of the product
$ab$ for given measurements $A_iB_j$. Here, $A_iB_j$ is often called the correlation function.
With these notions, one can further define the following Bell operator,
\be
S=A_0B_0+A_1B_0+A_0B_1-A_1B_1\label{bello}.
\ee
According to Eq. (\ref{bell3}) and some further calculations, one can find that
\be
\braket{A_0B_0}+\braket{A_1B_0}+\braket{A_0B_1}-\braket{A_1B_1}\leq 2,\label{cc}
\ee which is known as the (classical) CHSH inequality.
It is noted that the discussion above is abstract and has nothing to do with how to realize the measurements or whether the systems are classical or quantum.
The only ingredient is the classical (local) correlations.

However,  let us consider the two subsystems measured by Alice and Bob, which are two qubits in the singlet state $\ket{\Psi^-}=\frac{1}{\sqrt{2}}(\ket{01}-\ket{10})$, where $\ket{0}$ and $\ket{1}$ are the eigenstates of $\sigma_z$ for the eigenvalues of $+1$ and $-1$.
Suppose that the $A_0$ and $A_1$ correspond to the measurements of spin in the orthogonal directions $e_0$ and $e_1$, respectively. Similarly, $B_0$ and $B_1$ correspond to the measurements in the directions $-\frac{1}{\sqrt{2}}(e_0+e_1)$ and $\frac{1}{\sqrt{2}}(-e_0+e_1)$. Then, we have
\be
\braket{A_0B_0}+\braket{A_1B_0}+\braket{A_0B_1}-\braket{A_1B_1}= 2\sqrt{2},\label{qc}
\ee
as $\braket{A_0B_0}=\braket{A_1B_0}=\braket{A_0B_1}=-\braket{A_1B_1}=\frac{1}{\sqrt{2}}$.
As one can see, the violation of the classical bound given in Eq. (\ref{cc}) reveals  the non-local character of quantum theory.

\section{The internal nonlocality in simulating {\cal PT}-symmetric systems}
Now, we have known that in the simulation paradigm, a $\cal PT$-symmetric (pseudo-Hermitian) system can be embedded into a global Hermitian one.
Oppositely,  only with a global Hermitian Hamiltonian at hand, how can we know whether a pseudo-Hermitian Hamiltonian is embedded inside? or such a global Hermitian Hamiltonian is composed by local Hermitian ones?
Moreover, can we have the figure of merit to test the reliability of simulation paradigm?

A natural way to answer these questions is to investigate the internal nonlocality of the global Hermiticity.
As Eq. (\ref{Ht1}) suggests, the Hermitian dilation Hamiltonian may bring non-local correlations to the subsystems.
Inspired by the CHSH scenario, here, we propose a CHSH-like discussion on the nonlocality for the global Hamiltonian.
Owing to the fact that the nonlocality does not directly come from the entangled states but rather from the Hermitian dilation, our formulation differs considerably from earlier studies in essence. To distinguish between our setting and the CHSH scenario on quantum states, we coined the term {\it internal nonlocality}.

\subsection{The Simulation picture}
Without loss of generality, let us take the two dimensional model given in Eq. (\ref{Ht}) as an example.
In the simulation picture, the Hermitian dilation Hamiltonian $\hat{H}$ is assumed to be shared by Alice
and Bob.
Suppose Alice makes two local measurements denoted as $A_0$ and $ A_1$; while Bob also makes two local
measurements, denoted $B_0$ and $B_1$. By adopting similar ideas to those used in the CHSH scenario, one can consider the correlation functions $B_iA_j$. The reason why $B_i$ comes before $A_j$ is that  Alice is assumed to be in charge of the second subsystem, which simulates the $\cal PT$-symmetric Hamiltonian.
 Due to the fact that they are actually ``measuring'' the global Hamiltonian, Alice and Bob only need to use local states to obtain the measurement results.
Let Alice have the local state $\{\ket{u_+}=u\ket{0}+ v\ket{1}\}$ for $A_0$ and
$\{\ket{u_-}=\overline{v}\ket{0}-\overline{u}\ket{1}\}$ for $A_1$;
while Bob has two local states $\{ \ket{0}$ and $\ket{1}\}$ for $B_0$ and $B_1$, respectively.
Hence the expectations of $B_iA_j$ can be calculated as follows:
\begin{eqnarray}
&&\braket{B_0A_0}=Tr(\ket{0}\bra{0} \otimes \ket{u_+}\bra{u_+})\hat{H},\label{e1}\\
&&\braket{B_1A_0}=Tr(\ket{1}\bra{1} \otimes \ket{u_+}\bra{u_+})\hat{H},\label{e2}\\
&&\braket{B_0A_1}=Tr(\ket{0}\bra{0} \otimes \ket{u_-}\bra{u_-})\hat{H},\label{e3}\\
&&\braket{B_1A_1}=Tr(\ket{1}\bra{1} \otimes \ket{u_-}\bra{u_-})\hat{H}.\label{e4}
\end{eqnarray}
Now, one can further consider the expectation value of the Bell operator:
\bn
&&\braket{B_0A_0}+\braket{B_0A_1}+\braket{B_1A_0}-\braket{B_1A_1}\nonumber \\
&&=Tr [\ket{0}\bra{0}\otimes I_2+\ket{1}\bra{1}\otimes (|u|^2-|v|^2)\sigma_z \nonumber\\
&&\hspace{0.4in}+\ket{1}\bra{1}\otimes 2(u\overline{v}\ket{0}\bra{1}+\overline{u}v\ket{1}\bra{0})
]\hat{H} \nonumber\\
&&=Tr [\ket{0}\bra{0}\otimes\Lambda +\ket{1}\bra{1}\otimes
2 (u\overline{v}\ket{0}\bra{1}+\overline{u}v\ket{1}\bra{0})\Lambda]\nonumber\\
&&=2E_0+(\overline{u}v+u\overline{v})\,\omega_0\cos\alpha\label{exs}.
\en
Here, for the last term shown in Eq. (\ref{exs}), one also knows
\be
|(\overline{u}v+u\overline{v})\,\omega_0\cos\alpha|\leq |2s\cos^2\alpha|,\label{pers}
\ee
where the identity holds if and only if $u=\pm v$.
When the $\cal PT$-symmetric Hamiltonian approaches the exceptional point, that is, $\alpha\rightarrow \frac{\pi}{2}$, Eq. (\ref{exs}) gives only the value $2E_0$. However, Eq. (\ref{exs}) can reach $2E_0\pm2s$
with $\alpha= 0$, when we have a Hermitian Hamiltonian.
 This means that the $\cal PT$-symmetric system gives the largest departure from a Hermitian system when it tends to be broken. The unbroken $\cal PT$-symmetry can thus be viewed as an intermediate case.

\subsection{The Classical and Local Hermitian pictures}
In addition to the simulation picture, we  study the same setting but with another two pictures, namely the classical and local Hermitian pictures, in order to give a clear illustration of the internal nonlocality.
Firstly, the classical picture here means that one just skips all the details of quantum mechanics but only considers a classical description of what Alice and Bob do.
The only thing we ask for is to have the picture be consistent with the simulation picture.
That is, we assume that one of the two observers, e.g. Alice, has a ``$\cal PT$-symmetric like'' subsystem and the joint measurements of Alice and Bob depict the characteristics of measuring the global Hamiltonian $\hat{H}$.
Indeed, such a consistency rule plays a key role in giving a classical picture. A natural consequence of this rule is to assume
the measurement results of $A_j$ are just $\lambda_\pm$, namely the eigenvalues of the $\cal PT$-symmetric
Hamiltonian $H$.
At the same time,  the results of $B_i$ should be $1$, such that the correlation functions $B_iA_j$  trivially give the eigenvalues of $\hat{H}$.
Now, $A_i$ and $B_i$ are determined, equivalently completing the classical picture.
Moreover, since Bob's results always give $1$, apparently the two observers' results and the corresponding probability distributions are independent.
Thus,  we do have a classical local picture.

Let us come back to calculate the expectation of the Bell operator.
As the results of $A_i$ are the eigenvalues $\lambda_\pm$ and the result of $B_i$ is $1$, we have
\bn
&&\braket{B_0A_0}+\braket{B_0A_1}+\braket{B_1A_0}-\braket{B_1A_1}\nonumber \\
&&=\int[ B_0(\nu)(A_0+A_1)(\nu)+B_1(\nu)(A_0-A_1)(\nu)]d\nu\nonumber\\
&&=\int[ (A_0+A_1)(\nu)+(A_0-A_1)(\nu)]d\nu\nonumber\\
&&=2E_0+\omega_0\,(p_+-p_-).\label{exc}
\en
where $p_\pm$ are the probabilities corresponding to the situations when the results of $A_0$ are $\lambda_\pm$.

Secondly, let us consider the local Hermitian picture.
Now, the randomness comes from the global Hamiltonian $\hat{H}'$, which is in  a tensor product form of two local Hermitian Hamiltonians.
To have this local Hermitian picture be consistent with the simulation,
one can assume that $\hat{H}'$ has the same eigenvalues as $\hat{H}$ and one of the local Hamiltonians has the same eigenvalues as $H$.
Hence, we have  $\hat{H}'=I\otimes H_h$, where
$H_h=\lambda_+\ket{s_+}\bra{s_+}+\lambda_-\ket{s_-}\bra{s_-}$ and $\ket{s_\pm}$ are two orthogonal states.
In contrast to $\hat{H}$, the form of $\hat{H}'$ implies that it does not have internal nonlocality.
It also implies that by distinguishing the isospectral global Hamiltonians $\hat{H}$ and $\hat{H}'$, one can distinguish
 a $\cal PT$-symmetric Hamiltonian $H$ from an isospectral Hermitian Hamiltonian $H_h$.

Again, by  substituting the $\hat{H}'$ in the local Hermitian picture to  Eqs. (\ref{e1}-\ref{e4}), the expectation of the Bell operator is
\begin{eqnarray}
&&\braket{B_0A_0}+\braket{B_1A_0}+\braket{B_0A_1}-\braket{B_1A_1}\nonumber\\
&&=Tr (I\otimes \ket{u_+}\bra{u_+})(I\otimes H_h)\\
&&+Tr [(\ket{0}\bra{0}-\ket{1}\bra{1})\otimes \ket{u_-}\bra{u_-} ](I\otimes H_h),\nonumber
\end{eqnarray}
which can be further reduced to
\be
2\braket{u_+|H_h|u_+}=2\lambda_+|\braket{u_+|s_+}|^2+2\lambda_-|\braket{u_+|s_-}|^2.
\ee
As $\lambda_\pm=E_0\pm \frac{\omega_0}{2}$, we can denote $p_\pm=|\braket{u_+|s_\pm}|^2$ and reach
\bn
2E_0+\omega_0\,(p_+-p_-). \label{exl}
\en
By comparing Eq. (\ref{exl}) with Eqs. (\ref{exs}) and (\ref{exc}), 
all the expectations in the three pictures contain two terms.
The common term  $2E_0$ is the sum of the two eigenvalues $\lambda_+$ and $\lambda_-$; while the other
one represents a deviation term.
This deviation term is the same for the classical and local Hermitian pictures, as both of them do not support the internal nonlocality.
Moreover, we also have
\be
|\omega_0(p_+-p_-)|=|2s(p_+-p_-)\cos\alpha|\leq|2s\cos\alpha|,\label{percl}
\ee
which means that these two pictures give a larger value of the upper bound than that obtained in the simulation picture.

\section{Discussions}
Here, we discuss the physical implications behind our results by contrasting them with those of the CHSH
scenario~\cite{CHSH}.
Even though the generalization of  the CHSH scenario to $\cal PT$-symmetric settings can be found in the literature~\cite{Japaridze_2017}, these approaches essentially differ from our discussions.
In the CHSH scenario, the two observers share some entangled states and perform local measurements to explore the correlations.
On the contrary, in our setting, the resource of correlations comes from the Hermitian dilation Hamiltonian rather than
states.

Moreover, in the CHSH scenario, the observers do perform several local measurements.
For example, Alice can measure the spin in the $e_0$ and $e_1$ directions.
However, in our scenario, Alice performs two ``local measurements'' with two orthogonal local states $\ket{u_+}$
and $\ket{u_-}$.
According to von Neumann's measurement theory, these two states can only represent one measurement rather than two.
Further more, our randomness and correlations come from the global Hamiltonian.
Hence, Alice and Bob can obtain ``measurement results'' simply by inputting different states, reaching a similar
effect to the measurements in the CHSH scenario.

The most significant distinction between our discussions and CHSH's is that our scenario is concretely
constructed and logically derived by a consistency rule, which reflects the natural ideas and requirements in
simulations of $\cal PT$-symmetric systems.
This explains why the measurement results are a posteriori,  determined by the consistency rule in the classical
picture; while in the CHSH case they are a priori known.
It also explains why the classical and local Hermitian pictures have the same bounds.
Both the pictures are constructed to be consistent with the simulation.
The classical picture gives a general and abstract description of Alice's and Bob's measurements as well as the correlations in simulation, from the perspective of locality.
The local Hermitian picture can be viewed as a quantum realization of this classical (local) description.
Hence the same upper bounds of the two pictures is reasonable.

It is also worth noting that the expectation of the Bell operator exists in a larger range
for the classical and local Hermitian cases,  rather than in the simulation case.
At first glance, the results are counter-intuitive  as the latter case possesses internal nonlocality.
Indeed, non-local correlations yield a  larger range for the upper bound in the CHSH scenario.
However, our scenario is based on the consistency rule utilizing $\hat{H}$ to simulate $H$ and the
measurements of $\hat{H}$ to simulate measurements of $H$.
The results of Eqs. (\ref{exs}, \ref{exc}, \ref{exl}) are all essentially
characterizing the average deviation from the mean value $2E_0$ in the measuring process.
Note that $\hat{H}$ correlates the subsystems and the internal nonlocality can therefore be viewed to impose some internal constraints on the system.
As a result, it is reasonable to have a smaller deviation term in the simulation picture.
Moreover, when approaching the exceptional point, an unbroken $\cal PT$-symmetric system shows the largest departure from Hermitian systems. The minimal deviation at the exceptional point is consistent with such an intuition.

It should also be noted that the results of this paper mainly focus on the two dimensional case in Eq. (\ref{Ht}). However, Eq. (\ref{Ht}) is a special case of Eq. (\ref{Ht1}).
Hence, the analogy of the Hermitian dilation Hamiltonians, as well as the isospectral property, implies that the classical picture
 can be generalized in general. That is, for a higher dimensional Hamiltonian in Eq. (\ref{Ht1}),
we can similarly assume Alice's results to be the eigenvalues and Bob's result always be $1$, establishing a classical picture. By choosing meaningful Bell operators, a discussion on internal nonlocality is natural in higher dimensional spaces.

Before the conclusion, we propose two potential applications.
First, our results provide a figure of merit to know and test the reliability of simulation.
Suppose we have a set of devices, which can produce the Hermitian dilation Hamiltonian and simulate a $\cal PT$-symmetric system.
One may wonder whether the device is reliable, or does it faithfully realize the simulation design.
Apparently, this question is closely related to whether the Hermitian dilation Hamiltonian is well prepared.
To see this, one may have Alice and Bob perform the joint correlation measurements, comparing the results to Eq. (\ref{pers}).
If the obtained value of upper bound is larger than that given in Eq. (\ref{pers}), then the device cannot produce the needed global Hamiltonian and cannot be used for simulation. Otherwise, it is likely to be reliable.

Moreover, our results can also help in the verification problem of a $\cal PT$-symmetric system.
Consider the following scenario.
Let Alice have a system, which is either a simulated $\cal PT$-symmetric or an isospectral Hermitian one.
Let Bob be in charge of another system, which either serves as an ancillary subsystem in simulation or
a completely independent Hermitian system.
Can they verify whether Alice's system is $\cal PT$-symmetric or Hermitian just by making measurement?
Note that the isospectral property prevents one from seeing the difference by simply reading out the eigenvalues.
Moreover, there exist infinitely many isospectral Hermitian Hamiltonians.
To this end, Alice and Bob can locally measure the global Hermitian Hamiltonian and evaluate the joint correlation measurements.
If the randomness comes from the  classical or local Hermitian pictures, they can obtain a large deviation from the mean value.
Thus, they know that the system is not simulated to be $\cal PT$-symmetric.

\section{Conclusion}
In summary, we propose an operational way to explore the internal properties of $\cal PT$-symmetric systems, as well as their Hermitian dilations, by
constructing a non-local scenario between Alice and Bob. It is illustrated how to construct correlation pictures based on some
concrete procedures such as simulation, proposing a different aspect of investigating nonlocality.
By performing local measurements, the resulting expectation values make it possible to extract the internal
nonlocality in the global Hermiticity.
The ranges in different pictures clearly show the departure of $\cal PT$-symmetric systems from classical and
Hermitian quantum systems, for which the latter two share the same bound.
The extremal property of the exceptional point is obtained in the simulation picture.
These results not only show the characteristics of the internal nonlocality but they also can have potential applications. In addition, despite focusing on the discussion of $\cal PT$-symmetric systems, it is possible to generalize our discussions to the simulation of other non-Hermitian systems.

\section*{Acknowledgement}
This work is partially supported by the National Natural Science Foundation of China (Grants No. 11901526, No. 12031004, and No. 61877054),
the Ministry of Science and Technology, Taiwan under Grant No. 109-2112-M-007-019-MY3, the Science Foundation of Zhejiang Sci-Tech University (Grant No. 19062117-Y), and the China Postdoctoral
Science Foundation (Grant No. 2020M680074).



\end{document}